\documentclass[pra,twocolumn,floatfix,superscriptaddress,longbibliography,notitlepage, nofootinbib]{revtex4-1}
\pdfoutput=1

\usepackage[utf8]{inputenc}
\usepackage{braket}
\usepackage{amsmath}
\DeclareMathOperator{\Tr}{Tr}
\usepackage{dcolumn}   % needed for some tables
\usepackage{bm}        % for math
\usepackage{amssymb}
\usepackage{mathtools}
\usepackage{tikz}
\usepackage{qcircuit}
\usepackage{appendix}
\usepackage{hyperref}
\usepackage{subcaption}
\usepackage{amsmath,amsfonts,amssymb}
\usepackage{mathtools}
\usepackage{thmtools,thm-restate}
\usepackage{algorithm}
\usepackage{mathrsfs}
\usepackage{pgfplots}
\usetikzlibrary{3d}

\usepackage{algpseudocode}

\newtheorem{definition}{Definition}

\newtheorem{lemma}{Lemma}

\newcommand{\revise}[1]{\textcolor{black}{#1}}
\newcommand{\xbf}{\textbf{x}}

\newcommand{\Ibb}{\mathbb{I}}

\newcommand{\Rbb}{\mathbb{R}}

\newcommand{\Mcal}{\mathcal{M}}

\usetikzlibrary{arrows.meta}

\def\BibTeX{{\rm B\kern-.05em{\sc i\kern-.025em b}\kern-.08em
    T\kern-.1667em\lower.7ex\hbox{E}\kern-.125emX}}

\begin{document}
\title{Quantum Advantage in Testing (Local) Convexity and Monotonicity of Function}
\author{Nhat A. Nghiem}
\affiliation{Department of Physics and Astronomy, State University of New York at Stony Brook, Stony Brook, NY 11794-3800, USA}
\affiliation{C. N. Yang Institute for Theoretical Physics, State University of New York at Stony Brook, Stony Brook, NY 11794-3840, USA}

\begin{abstract}
   It is shown that a quantum computer can test the convexity and monotonicity of a given function exponentially more efficiently than a classical computer. This establishes another prominent example that showcases the potential of quantum computers in function-related problems, which can be practical in functional optimization. 
\end{abstract}
\maketitle

\section{Introduction}
Quantum computers hold great promise for solving problems that lie beyond the reach of classical computers. The intrinsic properties of quantum physics, such as entanglement and superposition, allow information to be stored and processed in a different manner, enabling advantage in solving certain computational problems. Numerous examples have been found, including quantum search algorithm \cite{grover1996fast}, factorization algorithm \cite{shor1999polynomial, regev2023efficient}, computing black-box \cite{deutsch1992rapid,deutsch1985quantum}, simulation algorithm \cite{lloyd1996universal, aharonov2006polynomial, berry2007efficient,berry2012black,berry2014high,berry2015hamiltonian, childs2010relationship, childs2022quantum, haah2021quantum, low2017optimal,low2019hamiltonian, an2022quantum, berry2020time,chen2021quantum,low2018hamiltonian,poulin2011quantum, kieferova2019simulating}, linear equation solver \cite{harrow2009quantum, childs2017quantum, subacsi2019quantum, nghiem2025new1, kerenidis2020quantum}, topological data analysis \cite{lloyd2016quantum, gyurik2022towards,berry2024analyzing, hayakawa2022quantum}, quantum machine learning algorithms \cite{lloyd2013quantum, lloyd2014quantum, wiebe2012quantum, wiebe2014quantum, mitarai2018quantum, schuld2014quest,schuld2018supervised,schuld2019quantum, schuld2020circuit}, etc.

Despite significant progress and there are certainly many more algorithms to be discovered, there is a major roadblock to the practical realization of quantum advantage. Many algorithms above, for example, quantum linear solver \cite{harrow2009quantum, childs2017quantum}, quantum supervised learning \cite{lloyd2013quantum}, quantum principal component analysis \cite{lloyd2014quantum}, quantum data fitting \cite{wiebe2012quantum}, assume a black-box/oracle in which a quantum computer can access classical data coherently. Quantum random access memory (QRAM) was proposed to realize this oracle \cite{giovannetti2008architectures,giovannetti2008quantum}, however, large-scale QRAM is still far away, thus deferring near-term realization of many quantum algorithms. More severely, in a series of seminal works \cite{tang2018quantum,tang2019quantum,tang2021quantum}, it was even shown that quantum advantage primarily comes from the black-box/oracle assumption. With an analogous assumption, a classical computer can tackle corresponding problems with at most polynomial slowdown, thus resisting many claimed exponential quantum speedups. Thereby, these results have raised an important question concerning whether a quantum computer can be advantageous without resorting to strong input assumptions. 

A few examples have been found with provable theoretical advantage. Bravyi et al. \cite{bravyi2018quantum} proved that the constant depth circuit can solve the problem involving binary quadratic form, whereas the classical circuit requires logarithmical depth. This advantage is maintained even in the presence of noise \cite{bravyi2020quantum}. Maslov et al. \cite{maslov2021quantum} showed that quantum scratch space is stronger than its classical counterpart. In \cite{gao2018quantum}, the authors introduced a quantum generative model and rigorously proved that it is stronger than the classical model. Liu et al. \cite{liu2021rigorous} constructed a supervised learning problem where the quantum classifier is provably more efficient than the classical classifier. Recently, it has been shown in \cite{nghiem2024simple} that it is possible to execute quantum gradient descent, which was first proposed in \cite{rebentrost2019quantum}, without oracle/black-box access. The idea and technique outlined in \cite{nghiem2024simple} have been carried out further in \cite{nghiem2025quantum1, nghiem2025quantum}, where the author showed that a wide range of problems, such as solving linear systems, nonlinear systems, constructing support vector machines and performing data fitting can also be efficiently handled by quantum algorithms without the need of oracle / black-box access.

\begin{figure}[t!]
    \centering
    \begin{center}
    \begin{tikzpicture}[scale = 2.0]
        % Draw Axes
        \draw[->] (-2,0) -- (2,0) node[right] {$x$};
        \draw[->] (0,-1.5) -- (0,2.5) node[above] {$y$};

        % Define the function
        \foreach \i in {0.95,1.04}
            \fill (\i,{2*\i^5 - 6*\i^3 + 4*\i}) circle (1pt) ;
        \foreach \i in {0.8,1.2}
            \fill (\i,{\i^3 - 2*\i + 1}) circle (1pt) ;
        
        \draw[blue, domain=-1.5:1.5, thick, samples=100, smooth,variable=\x] 
            plot (\x, {2*\x^5 - 6*\x^3 + 4*\x}) ;
        \draw[red, domain=-1.5:1.5, thick, samples=100, smooth,variable=\x] 
            plot (\x, {\x^3 - 2*\x + 1}) ;
        \draw[thick, dashed] (1, 0) circle (0.4);
        \fill[blue!20, opacity=0.3] (1,0) circle (0.4);
        %\draw[->] (1.2,2.0) -- (1.2, 0.5) ;
        \node at (1.3,2.4) {Key motivation:};
        \node at (1.3,2.2) {How does the function};
        \node at (1.3, 2.0) {look like in shaded region?};
        \node at (1.2, -1.0) {Quantum computer can test} ;
        \node at (1.2, -1.2) {sign of first, second derivative,} ;
        \node at (1.2, -1.4) {at chose points in parallel };
        \node at (1.2, -1.6) {and dissect the local shape, }; 
        \node at (1.2, -1.8) {achieving exponential enhancement};
        \node at (1.2, -2.0) {compared to classical counterpart};
        \node (A) at (1.2, -0.8) {};
        \node (B) at (0.95, 2*0.95^5 - 6*0.95^3 + 4*0.95) {}; 
        \draw[->] (B) to[out = 0, in = 30 ,looseness = 1.5] (A);
        
        \node (C) at (1.04, 2*1.04^5 - 6*1.04^3 + 4*1.04) {};
        \draw[->] (C) to[out = 0, in = 30 ,looseness = 1.5] (A);
        \node (D) at (0.8, 0.8^3 - 2*0.8 + 1) {};
        \draw[->] (D) to[out = 180, in = 150 ,looseness = 1.5] (A);

        \node (E) at (1.2, 1.2^3 - 2*1.2 + 1) {}; 
        \draw[->] (E) to[out = 0, in = 0 ,looseness = 1.5] (A);
    \end{tikzpicture}
\end{center}
    \caption{A plot of $y = 2x^5 - 6x^3 + 4x$ and $y=x^3 - 2x + 1$. These two functions exhibit complicated landscape across the whole $x-y$ domain. However, one may be interested in its local behavior, for example, its ``shape'' inside the shaded region as indicated above. Apparently from the plot, one can see that the function $y=x^3 - 2x + 1$ (with red color) is convex inside this region, and the blue colored function is monotonically decreasing.  }
    \label{fig: function}
\end{figure}
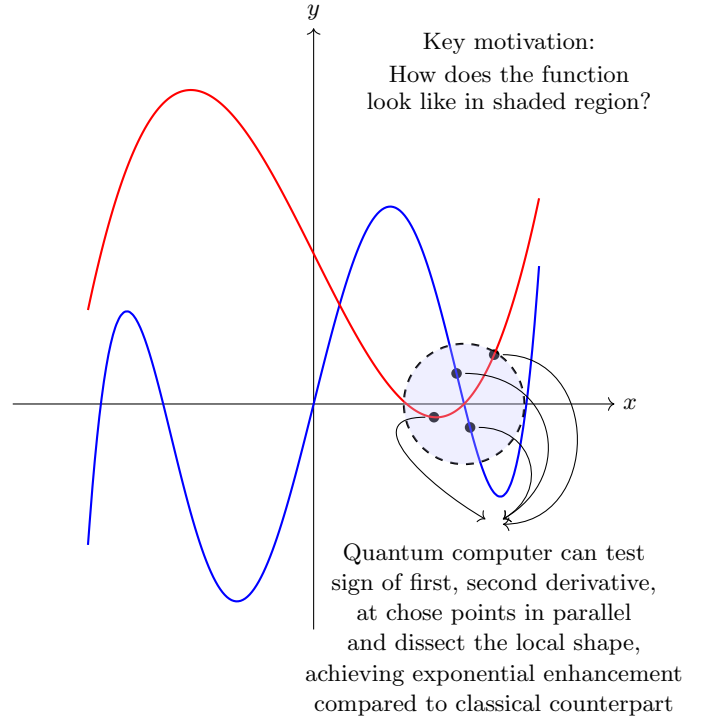

In this work, motivated by the aforementioned line of research, we explore how quantum computers perform in function-related problems. Specifically, we focus on two key aspects: convexity and monotonicity, which characterize the ``shape" of the given function within a certain domain (see Figure \ref{fig: function} for illustration). To examine convexity, we develop three quantum algorithms based on the first derivative test, the second derivative test, and Jensen's inequality. Broadly speaking, the first derivative test leverages the fact that the first-order derivative (if it exists) of a (locally) convex function is nondecreasing within corresponding domain. The second derivative test relies on the sign of the second-order derivative (if it exists), which should hold positive for a convex function. Both approaches, however, apply only to univariate polynomials. For multivariate polynomials, a more general method involves Jensen's inequality, which assesses the function's values at multiple points to determine whether they satisfy a specific inequality. We demonstrate that the ability of quantum computers to store and process classical information (data points) using logarithmic resources, combined with recent advances in quantum computation \cite{nghiem2024improved, nghiem2023improved1, gilyen2019quantum}, enables us to analyze the behavior of first- and second-order derivatives in (poly)logarithmic time. As a result, we reveal convexity efficiently, achieving an exponential speedup compared to classical algorithms. Similarly, the third approach, based on Jensen's inequality, examines the function's behavior across multiple points. The ability of quantum computers to process these points simultaneously facilitates testing whether Jensen's inequality holds, thereby confirming convexity. Furthermore, the techniques we develop for convexity testing can be naturally extended to monotonicity testing with only minor modifications, leading to an exponential speedup in this setting as well. We point out that a quantum algorithm for convexity testing has recently appeared in \cite{nghiem2024quantum1}. However, their method requires an oracle/black-box assumption and is only applicable to some specific types of polynomial. As we pointed out before, this assumption is not completely justified and, as we will see subsequently, our method can work for a broader range of polynomials. 

Before outlining our proposal in detail, we remark that many of the important recipes that appear in our subsequent discussion are provided in the Appendix \ref{sec: prelim}. Thus, we strongly encourage the readers to take a look at these preliminaries and then return to the main text.

%\section{Testing Convexity}
%\label{sec: testconvexity}
%\subsection{Overview }
%\label{sec: overview}

\section{Testing convexity of univariate polynomial}
\label{sec: testingunivariate}
Consider a univariate polynomial $f(x):\Rbb \longrightarrow \Rbb$ and examine the shape of such a function in some domain $\mathcal{D}$. We note that by trivially redefining the function, it is always possible to choose $\mathscr{D} = [-\frac{1}{2}, \frac{1}{2}]$, therefore, through the remaining, we work with this domain for simplicity. Without loss of generalization, assume that $|f(\xbf)| \leq \frac{1}{2}$ for all $x \in \mathscr{D}$, and also  $|\frac{\partial f(x)}{\partial x}| \leq \mathcal{P}$, $|\frac{\partial^2 f(x)}{\partial^2 x}| \leq \mathcal{Q}$. For the purpose of testing, we choose $n$ points $x_1,x_2,...,x_n \in \mathscr{D}$. Define $n$-dimensional vector $\xbf = (x_1,x_2,...,x_n)^T$. For a purpose that would be clear later, we first construct a block encoding of $\rm diag (\xbf)$. Given that these points are classically known, we can use any of the amplitude encoding methods \cite{grover2000synthesis,grover2002creating,plesch2011quantum, schuld2018supervised, nakaji2022approximate,marin2023quantum,zoufal2019quantum, zhang2022quantum} to construct the state $\frac{\xbf}{||\xbf||}$, using a circuit of depth $\mathcal{O}(\log n)$. Then we leverage the result of \cite{rattew2023non, guo2024nonlinear} (see Lemma \ref{lemma: diagonal} in the appendix \ref{sec: prelim}) to use this state and construct the block encoding of $\rm diag (\xbf)/||\xbf||$, incurring further circuit depth $\mathcal{O}(\log n)$. The factor $||\xbf||$ can be removed using Lemma \ref{lemma: amp_amp}. Thus, we obtain the block encoding of $\rm diag (\xbf)$ in complexity $\mathcal{O}\big(  \log n\big)$.
  
For the next step, we need the following essential result from \cite{gilyen2019quantum}:
\begin{lemma}\label{lemma: qsvt}[\cite{gilyen2019quantum} Theorem 56]
\label{lemma: theorem56}  
Suppose that $U$ is an
$(\alpha, a, \epsilon)$-encoding of a Hermitian matrix $A$. (See Definition 43 of~\cite{gilyen2019quantum} for the definition.)
If $P \in \mathbb{R}[x]$ is a degree-$d$ polynomial satisfying that
\begin{itemize}
\item for all $x \in [-1,1]$: $|P(x)| \leq \frac{1}{2}$,
\end{itemize}
then, there is a quantum circuit $\tilde{U}$, which is an $(1,a+2,4d \sqrt{\frac{\epsilon}{\alpha}})$-encoding of $P(A/\alpha)$ and
consists of $d$ applications of $U$ and $U^\dagger$ gates, a single application of controlled-$U$ and $\mathcal{O}((a+1)d)$
other one- and two-qubit gates.
\end{lemma}
The above lemma allows us to transform the block-encoded operator $\rm diag(\xbf)$ into $\Mcal =  \sum_{i=1}^n f(x_i) \ket{i-1}\bra{i-1} $. Let the degree of $f(x)$ be $\deg(f)$, then the complexity of this step is $\mathcal{O}\big( \deg (f) \log n\big)$. We remark that as $f(x)$ is a polynomial, its first derivative and second derivative are also polynomials (of degree $\deg (f)-1,\deg(f) -2$, respectively). Thus, we can also use the above lemma to construct the block encoding of 
\begin{align}
    \mathcal{M}_1 =  \frac{1}{\mathcal{P}} \sum_{i=1}^n f'(x_i) \ket{i-1}\bra{i-1}, \\  \mathcal{M}_2 = \frac{1}{\mathcal{Q}} \sum_{i=1}^n f''(x_i) \ket{i-1}\bra{i-1} 
\end{align}
in complexity $\mathcal{O}\big( (\deg(f)-1) \log n \big)$ and $\mathcal{O}\big( (\deg(f)-2) \log n \big)$, respectively. Using these results, we shall outline our quantum convexity testing algorithm in the following. 

\subsubsection{Approach based on second derivative test}
\label{sec: secondderivativetest}
This approach works under the condition that $f(x)$ should be twice-differentiable. Thus, it applies only when $f(x)$ contains a power of at least two. Given this, the function $f(x)$ is convex on $ [-\frac{1}{2}, \frac{1}{2}]$ if $f''(x) \geq 0$ for all $x \in  [-\frac{1}{2}, \frac{1}{2}]$.  In order to dissect the convexity, we choose $n$ points $x_1,x_2,...,x_n$ as above. Recall from above that we have the block encoding of $\mathcal{M}_2$. This matrix is diagonal, and its minimum eigenvalue, denoted as $\lambda_{\min}(\mathcal{M}_2)$ is also $\min \{ \frac{1}{Q} f''(x_1) , \frac{1}{Q} f''(x_2) ,...,  \frac{1}{Q} f''(x_n)   \}$. So our strategy is to look at the minimum eigenvalue of $\mathcal{M}_2$. If it is greater than zero, then it indicates that all remaining ones are also greater than zero, implying that $f''(x) > 0$ for all $x \in \{x_1,x_2,...,x_n\}$. 

As the next step, we use the block encoding of $\mathcal{M}_2$ combined with Lemma \ref{lemma: sumencoding} to construct the block encoding of $ \frac{1}{2}\Big( \Ibb_n - \mathcal{M}_2\Big)  $. The reason for this step is to shift the spectrum, so that the eigenvalues of the resultant matrix fall between $(0,1)$, indicating that the matrix is positive-semidefinite. The maximum eigenvalue of the resultant matrix turns out to be $\frac{1}{2} \big(1- \lambda_{\min}(\mathcal{M}_2) \big)$. If $\lambda_{\min}(\mathcal{M}_2)$ is greater than zero, it means that $ \frac{1}{2} \big(1- \lambda_{\min}(\mathcal{M}_2) \big) $ is smaller than $\frac{1}{2}$, and vice versa. The following result of \cite{nghiem2022quantum, nghiem2024improved, nghiem2023improved} allows us to estimate the largest eigenvalue of a positive matrix: 
\begin{lemma}
\label{lemma: largestsmallest}
    Given the block encoding of a positive-semidefinite Hermitian matrix $A$ of size $n\times n$ (assumed to have $\mathcal{O}(1)$ gap between two largest eigenvalues), the largest eigenvalue can be estimated up to additive accuracy $\epsilon$ in complexity $\mathcal{O}\Big(  T_A \frac{1}{\epsilon} \big(\log n +  \log \frac{1}{\epsilon}\big)\Big)$ where $T_A$ is the complexity of producing block encoding of $A$. 
\end{lemma}
Thus, setting $\frac{1}{2}$ as a threshold, a direct application of the above lemma can reveal the sign of $ \frac{1}{2}\big( 1- \lambda_{\min}(\Mcal_2)\big)$, which in turn reveals the sign of $\lambda_{\min}(\mathcal{M}_2)$, which can then be used to infer the convexity landscape of $f(x)$. We recall that the complexity of obtaining the block encoding of $\Mcal_2$ is $\mathcal{O}\Big( (\deg(f) -2) \log n\Big) $, so the complexity after using the above lemma is $\mathcal{O}\left((\deg(f) -2) \log (n) \frac{1}{\epsilon} \big(\log n +  \log \frac{1}{\epsilon}\big)\right)$. Assume that $\frac{1}{\epsilon}, \deg(f) \in \mathcal{O}(1)$, we have the comlexity for dissecting convexity using second derivative test is $\mathcal{O}\big(  \log^2 n \big)$. 

\subsubsection{Approach based on first derivative test}
In the case where the second derivative does not exist, the first derivative test can be used instead. It states that a function is convex if its first derivative is non-decreasing: $f'(x_2) \geq f'(x_1)$ for all $x_1 < x_2$ within $\mathscr{D}$. Apparently, if we use the same strategy as before, with $\Mcal_1$ instead of $\Mcal_2$ and impose the order $x_1<x_2 < ... <x_n$, it is not going to work because the sign of minimum eigenvalue of $\Mcal_1$ does not directly imply that $f'(x_1) \geq f'(x_2) \geq ... \geq f'(x_n) $. However, such a strategy can work with a slight modification. If we can somehow construct the block encoding of $\Mcal_3=  \frac{1}{\mathcal{P}}\sum_{i=1}^n \big(f'(x_{i+1})-f'(x_i) \big)\ket{i-1}\bra{i-1} $ with a newly defined term $f'(x_{n+1}) \equiv f'(x_1)$, then the minimum eigenvalue of such a matrix, denoted as $\lambda_{\min}(\Mcal_3)$, is exactly $\min \{ f'(x_{i+1})-f'(x_i)   \}_{i=1}^n $. If it is greater than zero, then it means that for all $i=1,2,...,n$, $ f'(x_{i+1})-f'(x_i) > 0 $, implying that the function is convex. Otherwise if if is smaller than zero, it means that there exist some $i \in [1,2,...,n]$ such that $ f'(x_{i+1})-f'(x_i)  < 0$, implying that the function is not convex in $\mathscr{D}$. As the procedure is rather lengthy and technical, we leave the construction in the Appendix \ref{sec: constructingm1}, and provide the main result in the following:
\begin{lemma}
    \label{lemma: constructingm1}
    There exists a quantum circuit of depth $\mathcal{O}\big( \deg(f)  \log n\big)$ that is a block encoding of  $\frac{1}{\sqrt{n}}\Mcal_3$. 
\end{lemma}
We note that as the identity matrix $\Ibb_n$ can be simply block-encoded (see below Def.~\ref{def: blockencode}), the block encoding of $\frac{1}{\sqrt{n}}\Ibb_n$ is easily constructed by using Lemma \ref{lemma: scale}. Given this, we can proceed similarly to the previous section, first building the block encoding of $\frac{1}{2\sqrt{n}}\big( \Ibb_n - \Mcal_3 \big)$, then applying lemma \ref{lemma: largestsmallest} to find out the sign of $\frac{1}{2\sqrt{n}} \big( 1- \lambda_{\min}(\Mcal_3) \big)$ (setting $\frac{1}{2\sqrt{n}}$ as a threshold), whereby inferring the sign of $\lambda_{\min}(\Mcal_3)$. It can finally be used to dissect the convexity of $f(x)$ in the domain $\mathscr{D}$. The complexity of this approach is the product of complexity of the above Lemma and Lemma  \ref{lemma: largestsmallest}, which is $\mathcal{O}\big(  \log^2 n \big)$, with the same premise that $ \frac{1}{\epsilon}, \deg(f) \in \mathcal{O}(1)$. 

\subsubsection{Approach based on Jensen's inequality}
This is the most general definition of convexity, where it states that for every finite collection of points $x_1,x_2,...,x_n \in \mathscr{D}$ and a non-negative reals $\lambda_1,\lambda_2,...,\lambda_n \geq 0$ satisfying $\sum_{i=1}^n \lambda_i = 1$ and $\sum_{i=1}^n \lambda_i x_i \in \mathscr{D}$, if $f \big( \sum_{i=1}^n \lambda_i x_i  \big) \leq \sum_{i=1}^n \lambda_i f(x_i)$, then $f(x)$ is convex. Our strategy is to estimate these quantities then compare directly. 

We recall from above that we have obtained the block encoding of $\rm diag (\xbf)$, denoted as $U_X$. Since $\lambda_1,\lambda_2,...,\lambda_n$ are known and their summation is one, the same amplitude encoding technique can be used to generate the state $\sum_{i=1}^n \sqrt{\lambda_i} \ket{i-1}$. For the purpose of presentation, we leave the details to the Appendix \ref{sec: detaillambdax}, in which we prove the following:
\begin{lemma}
\label{lemma: detaillambdax}
   Given the block encoding of $\rm diag (\xbf)$ and unitary that generates the state $\sum_{i=1}^n \sqrt{\lambda_i} \ket{i-1} $, both of depth $\mathcal{O}( \log n)$. There is a quantum circuit of depth $\mathcal{O}(\log n)$ that prepares the block encoding of $\big( \sum_{i=1}^n \lambda_i x_i   \big) \ket{0}\bra{0} - \big( \sum_{i=1}^n \lambda_i x_i \big) \ket{1}\bra{1} $
\end{lemma}
From the above block encoding, we can use Lemma \ref{lemma: theorem56} to transform it into $ f\big( \sum_{i=1}^n \lambda_i x_i   \big) \ket{0}\bra{0} - f\big( \sum_{i=1}^n \lambda_i x_i \big) \ket{1}\bra{1} $. Using such the unitary block encoding and apply it to the state $\ket{\bf 0} \ket{0}$, then according to Definition \ref{def: blockencode} and Eqn.~\ref{eqn: action}, we obtain the state:
\begin{align}
    \ket{\Phi} = \ket{\bf 0} f\big( \sum_{i=1}^n \lambda_i x_i   \big) \ket{0} + \ket{\rm Garbage}
\end{align}
By using amplitude estimation \cite{manzano2023real, rall2023amplitude, rall2021faster,brassard1997quantum}, we can estimate the amplitude of $ \ket{\bf 0}\ket{0}$, which is $f\big( \sum_{i=1}^n \lambda_i x_i   \big) $. 

In fact, using the same method as the above lemma, by replacing $\rm diag(\xbf)$ with the block encoding of $\Mcal =  \sum_{i=1}^n f(x_i) \ket{i-1}\bra{i-1}$ which we constructed earlier, we can produce the block encoding of $\sum_{i=1}^n \lambda_i f(x_i) \ket{0}\bra{0} - \sum_{i=1}^n \lambda_i f(x_i) \ket{1}\bra{1} $. Using the same procedure as above, applying such unitary to the state $\ket{\bf 0}\ket{0}$, and use amplitude estimation, we can estimate the value of $\sum_{i=1}^n \lambda_i f(x_i)  $. The complexity of this procedure is $\mathcal{O}(\log n)$, with more details will be provided in the Appendix \ref{sec: detaillambdax}. 

With the estimations of $f\big( \sum_{i=1}^n \lambda_i x_i   \big) $ and $\sum_{i=1}^n \lambda_i f(x_i)  $, a direct comparison can be made, which reveals the convexity of $f(x)$, according to Jensen's inequality. This approach achieves $\mathcal{O}(\log n)$ complexity -- which is quadratically more efficient than the previous two approaches using the first and second derivative tests.

\section{Testing convexity of multivariate polynomial}
\label{sec: testingmultivariate}
The above results have motivated us to go beyond the single-variable regime, and consider whether the quantum advantage still persists in the multivariate regime. This case exhibits more complication due to more variables, which resists the first and second derivative test. A more popular criterion that can work with any type of function is the positive-semidefiniteness of Hessian, a matrix that contains the second-order partial derivative of given function with respect to all variables. However, constructing Hessian for a general function is quite tricky, at least it is not within the reach of the technique provided in this work. Fortunately, the third approach, which is based on Jensen's inequality, can be naturally extended to a multivariate setting. More specifically, in the new domain $\mathscr{D} = [-\frac{1}{2}, \frac{1}{2}]^d$ (where $d >1$ is the dimension), a function $f(\xbf)$ is convex if for every finite collection of points $\xbf_1,\xbf_2,...,\xbf_n \in \mathscr{D}$ and a non-negative reals $\lambda_1,\lambda_2,...,\lambda_n \geq 0$ satisfying $\sum_{i=1}^n \lambda_i = 1$ and $\sum_{i=1}^n \lambda_i x_i \in \mathscr{D}$, if $f \big( \sum_{i=1}^n \lambda_i \xbf_i  \big) \leq \sum_{i=1}^n \lambda_i f(\xbf_i)$. 

Without loss of generalization, assume that $f(\xbf) = \sum_{k=1}^K f_k(\xbf) = \sum_{k=1}^K a_k x_1^{k_1} x_2^{k_2} ... x_d^{k_d}$ for $k_1,k_2,...,k_d, K \in \mathbb{Z}$, and that $|f(\xbf)| \leq \frac{1}{2} \forall x \in \mathscr{D}$. Additionally, the norm of its gradient, $\big| \bigtriangledown f(\xbf) \big|$ is bounded by $P$. Define $n$ points $\in \mathscr{D}$ as follows $\xbf_1 = (x_{1,1}, x_{2,1}, ..., x_{d,1})^T, \xbf_2 =  (x_{1,2}, x_{2,2}, ..., x_{d,2})^T$, ..., $\xbf_n =  (x_{1,n}, x_{2,n}, ..., x_{d,n})^T$. Earlier, at the beginning of Section \ref{sec: testingunivariate}, we discussed a procedure for producing the block encoding of $\rm diag (\xbf)$ for a $n$-dimensional vector $\xbf$ with efficient state preparation. Using the same procedure, we construct the block encoding of 
\begin{align}
\label{eqn: diag}
    \rm diag \big( x_{1,1},x_{1,2},...,x_{1,n}  \big)^T, \\ \rm diag \big( x_{2,1},x_{2,2},...,x_{2,n}  \big)^T, ..., \\ \rm diag \big(  x_{d,1},x_{d,2},..., x_{d,n} \big)^T
\end{align}
All with complexity $\mathcal{O}(\log n)$. The following recipe, which is the result of \cite{nghiem2025quantum}, is central to our subsequent construction
\begin{lemma}
    \label{lemma: multivariatefunction}
 Given the block encoding of operators defined in Eqn.~\ref{eqn: diag}. Assume that $k_1,k_2,...,k_d, K \in \mathcal{O}(1)$.   The block encoding of the diagonal matrix $\sum_{i=1}^n  f(\xbf_i) \ket{i-1}\bra{i-1}$ can be constructed using a quantum circuit of depth $\mathcal{O}\big(  \log n \big) $
\end{lemma}
For completeness, we provide the proof of the above lemma, which is the procedure outlined in \cite{nghiem2025quantum} in the Appendix \ref{sec: multivariatefunction}. Using the result of the above lemma combined with Lemma \ref{lemma: detaillambdax}, we can construct the block encoding of $ \big( \sum_{i=1}^n  \lambda_if(\xbf_i) \big) \ket{0}\bra{0} - \big( \sum_{i=1}^n  \lambda_if(\xbf_i) \big) \ket{1}\bra{1} $, which can be used to provide an estimation of $ \big( \sum_{i=1}^n  \lambda_if(\xbf_i) \big) $, according to the procedure below Lemma \ref{lemma: detaillambdax}. 

In order to estimate $f \big( \sum_{i=1}^n \lambda_i \xbf_i  \big)$, we need a slight modification of the method underlying the above lemma. More specifically, we first use Lemma \ref{lemma: detaillambdax} to construct the block encoding of 
$$ \big(  \sum_{i=1}^n \lambda_i x_{j,i} \big) \ket{0}\bra{0} -\big(  \sum_{i=1}^n \lambda_i x_{j,i} \big)\ket{1}\bra{1}  $$ 
for $j=1,2,..., d$. Note that the above operator is $\rm diag \big(  \sum_{i=1}^n \lambda_i x_{j,i}, -  \sum_{i=1}^n \lambda_i x_{j,i}   \big)^T$. Then we use the above lemma to construct the block encoding of $ f\big( \sum_{i=1}^n \lambda_i \xbf_i \big) \ket{0}\bra{0} - f\big( \sum_{i=1}^n \lambda_i \xbf_i \big)\ket{1}\bra{1}  $, which can be used to estimate $f\big( \sum_{i=1}^n \lambda_i \xbf_i \big)$ with the procedure we outlined earlier (below Lemma \ref{lemma: detaillambdax}). As both Lemma \ref{lemma: detaillambdax} and Lemma \ref{lemma: multivariatefunction} has complexity $\mathcal{O}(\log n)$, this approach has the total complexity also $\mathcal{O}(\log n)$.

\section{Testing Monotonicity}
%\label{sec: discussionandextension}
The above results have motivated us to go beyond convexity and consider other properties of analytical function. A particular property that can also (locally) capture the shape of given function is monotonicity. A univariate function is called monotonically increasing within the domain $\mathscr{D}$ if, for example, there are $n$ points with order  $x_1 < x_2 < ... < x_n $, we have $f(x_1) < f(x_2) < ... < f(x_n)$. A simple way to deduce the monotonicity is based on the first derivative: for all $x \in \mathscr{D}$, $f'(x) \geq 0$. Recall from previous discussion that we have the block encoding of:
\begin{align}
     \mathcal{M}_1 =  \frac{1}{\mathcal{P}} \sum_{i=1}^n f'(x_i) \ket{i-1}\bra{i-1}
\end{align}
We remind from Section \ref{sec: secondderivativetest} that we outlined a procedure using the block encoding of $\mathcal{M}_2 = \frac{1}{\mathcal{Q}} \sum_{i=1}^n f''(x_i) \ket{i-1}\bra{i-1}  $ to reveal the convexity of $f(x)$. The exact same procedure can be used to dissect whether $f'(x)$ is $\geq 0$ within the domain $\mathscr{D}$, by selecting $n$ different points and test the sign of the derivative of $f(x)$ at these points. The complexity for this procedure is $\mathcal{O}(\log^2 n)$, which is exponentially better than classical approach. We remark that the definition of monotonically decreasing is essentially the same, except that the order is reversed. Therefore, the strategy outlined above can be adapted to test whether $f(x)$ is monotonically decreasing as well. Thus, a quantum computer can test the monotonicity exponentially more efficient than classical counterpart, providing another instance, beside the convexity testing, that demonstrate quantum advantage. 

\section{Outlook and Conclusion}
In this work, we have successfully shown that quantum computers can test the convexity and monotonicity of a given function exponentially better than their classical counterparts. Our algorithms leverage a few techniques from the context, such as block encoding and quantum eigenvalue finding, combined with the insight from the derivative tests plus Jensen's inequality. Moreover, our work does not assume any sort of oracle/black-box procedure, thus clearly indicating the provable theoretical advantage of our results. It adds another instance to the existing literature, including \cite{bravyi2018quantum,bravyi2020quantum, maslov2021quantum, nghiem2025quantum,nghiem2025quantum1}, demonstrating the potential of quantum computers. Our results have provided great motivation for exploring quantum computational advantage toward problems involving analytical function, which is a topic of highly mathematical, yet potentially applicable to many areas. For example, convexity is critical in the context of convex optimization, and provided that one can dissect the convexity of a given function within some domain, can be very useful for initialization as well as application of optimization method, e.g., gradient descent. What kind of application in which our results can prove useful is a fascinating challenge, and we leave it for future works.

\section*{Acknowledgement}
 We acknowledge support from Center for Distributed Quantum Processing, Stony Brook University.

\bibliography{ref.bib}{}
\bibliographystyle{unsrt}

%\clearpage
%\newpage
\onecolumngrid
\appendix
\section{Preliminaries}
\label{sec: prelim}
Here, we summarize the main recipes of our work, which mostly derived in the seminal QSVT work \cite{gilyen2019quantum}. We keep the statements brief and precise for simplicity, with their proofs/ constructions referred to in their original works.

\begin{definition}[Block Encoding Unitary]~\cite{low2017optimal, low2019hamiltonian, gilyen2019quantum}
\label{def: blockencode} 
Let $A$ be some Hermitian matrix of size $N \times N$ whose matrix norm $|A| < 1$. Let a unitary $U$ have the following form:
\begin{align*}
    U = \begin{pmatrix}
       A & \cdot \\
       \cdot & \cdot \\
    \end{pmatrix}.
\end{align*}
Then $U$ is said to be an exact block encoding of matrix $A$. Equivalently, we can write $U = \ket{ \bf{0}}\bra{ \bf{0}} \otimes A + (\cdots)$, where $\ket{\bf 0}$ refers to the ancilla system required for the block encoding purpose. In the case where the $U$ has the form $ U  =  \ket{ \bf{0}}\bra{ \bf{0}} \otimes \Tilde{A} + (\cdots) $, where $|| \Tilde{A} - A || \leq \epsilon$ (with $||.||$ being the matrix norm), then $U$ is said to be an $\epsilon$-approximated block encoding of $A$. Furthermore, the action of $U$ on some quantum state $\ket{\bf 0}\ket{\phi}$ is:
\begin{align}
    \label{eqn: action}
    U \ket{\bf 0}\ket{\phi} = \ket{\bf 0} A\ket{\phi} +  \ket{\rm Garbage},
\end{align}
where $\ket{\rm Garbage }$ is a redundant state that is orthogonal to $\ket{\bf 0} A\ket{\phi}$. The above definition has multiple natural corollaries. 

\noindent
\textbf{Corollaries.} 
\begin{itemize}
    \item First, an arbitrary unitary $U$ block encodes itself
    \item Second, suppose that $A$ is block encoded by some matrix $U$, then $A$ can be block encoded in a larger matrix by simply adding any ancilla (supposed to have dimension $m$), then note that $\Ibb_m \otimes U$ contains $A$ in the top-left corner, which is block encoding of $A$ again by definition 
    \item Third, it is almost trivial to block encode identity matrix of any dimension. For instance, we consider $\sigma_z \otimes \Ibb_m$ (for any $m$), which contains $\Ibb_m$ in the top-left corner. 
\end{itemize}
\end{definition}

\begin{lemma}[\cite{gilyen2019quantum} \revise{Block Encoding of a Density Matrix}]
\label{lemma: improveddme}
Let $\rho = \Tr_A \ket{\Phi}\bra{\Phi}$, where $\rho \in \mathbb{H}_B$, $\ket{\Phi} \in  \mathbb{H}_A \otimes \mathbb{H}_B$. Given unitary $U$ that generates $\ket{\Phi}$ from $\ket{\bf 0}_A \otimes \ket{\bf 0}_B$, then there exists a highly efficient procedure that constructs an exact unitary block encoding of $\rho$ using $U$ and $U^\dagger$ a single time, respectively.
\end{lemma}

The proof of the above lemma is given in \cite{gilyen2019quantum} (see their Lemma 45). \\

% \begin{lemma}[Block Encoding of Product of Two Matrices]
% \label{lemma: product}
%     Given the unitary block encoding of two matrices $A_1$ and $A_2$, then there exists an efficient procedure that constructs a unitary block encoding of $A_1 A_2$ using each block encoding of $A_1,A_2$ one time. 
% \end{lemma}

% The proof of the above lemma is also given in \cite{gilyen2019quantum}, see their Lemma 53). 

\begin{lemma}[Block Encoding of Product of Two Matrices]
\label{lemma: product}
    Given the unitary block encoding of two matrices $A_1$ and $A_2$, then there exists an efficient procedure that constructs a unitary block encoding of $A_1 A_2$ using each block encoding of $A_1,A_2$ one time. 
\end{lemma}

\begin{lemma}[\cite{camps2020approximate} \revise{Block Encoding of a Tensor Product}]
\label{lemma: tensorproduct}
    Given the unitary block encoding $\{U_i\}_{i=1}^m$ of multiple operators $\{M_i\}_{i=1}^m$ (assumed to be exact encoding), then, there is a procedure that produces the unitary block encoding operator of $\bigotimes_{i=1}^m M_i$, which requires \revise{parallel single uses} of 
    %each 
    $\{U_i\}_{i=1}^m$ and $\mathcal{O}(1)$ SWAP gates. 
\end{lemma}
The above lemma is a result in \cite{camps2020approximate}. 
\begin{lemma}[\revise{\cite{gilyen2019quantum} Block Encoding of a  Matrix}]
\label{lemma: As}
    Given oracle access to $s$-sparse matrix $A$ of dimension $n\times n$, then an $\epsilon$-approximated unitary block encoding of $A/s$ can be prepared with gate/time complexity $\mathcal{O}\Big(\log n + \log^{2.5}(\frac{s^2}{\epsilon})\Big).$
\end{lemma}
This is presented in~\cite{gilyen2019quantum} (see their Lemma 48), and one can also find a review of the construction in~\cite{childs2017lecture}. We remark further that the scaling factor $s$ in the above lemma can be reduced by the preamplification method with further complexity $\mathcal{O}({s})$~\cite{gilyen2019quantum}.

\begin{lemma}[\cite{gilyen2019quantum} Linear combination of block-encoded matrices]
    Given unitary block encoding of multiple operators $\{M_i\}_{i=1}^m$. Then, there is a procedure that produces a unitary block encoding operator of \,$\sum_{i=1}^m \pm M_i/m $ in complexity $\mathcal{O}(m)$, e.g., using block encoding of each operator $M_i$ a single time. 
    \label{lemma: sumencoding}
\end{lemma}

\begin{lemma}[Scaling Block encoding] 
\label{lemma: scale}
    Given a block encoding of some matrix $A$ (as in~\ref{def: blockencode}), then the block encoding of $A/p$ where $p > 1$ can be prepared with an extra $\mathcal{O}(1)$ cost.  
\end{lemma}
To show this, we note that the matrix representation of RY rotational gate is
\begin{align}
   R_Y(\theta) = \begin{pmatrix}
        \cos(\theta/2) & -\sin(\theta/2) \\
        \sin(\theta/2) & \cos(\theta/2) 
    \end{pmatrix}.
\end{align}
If we choose $\theta$ such that $\cos(\theta/2) = 1/p$, then Lemma~\ref{lemma: tensorproduct} allows us to construct block encoding of $R_Y(\theta) \otimes \mathbb{I}_{{\rm dim}(A)}$  (${\rm dim}(A)$ refers to dimension of matirx $A$), which contains the diagonal matrix of size ${\rm dim}(A) \times {\rm dim}(A)$ with entries $1/p$. Then Lemma~\ref{lemma: product} can construct block encoding of $(1/p) \ \mathbb{I}_{{\rm dim}(A)} \cdot A = A/p$.  \\

The following is called amplification technique:
\begin{lemma}[\cite{gilyen2019quantum} Theorem 30; \revise{\bf Amplification}]\label{lemma: amp_amp}
Let $U$, $\Pi$, $\widetilde{\Pi} \in {\rm End}(\mathcal{H}_U)$ be linear operators on $\mathcal{H}_U$ such that $U$ is a unitary, and $\Pi$, $\widetilde{\Pi}$ are orthogonal projectors. 
Let $\gamma>1$ and $\delta,\epsilon \in (0,\frac{1}{2})$. 
Suppose that $\widetilde{\Pi}U\Pi=W \Sigma V^\dagger=\sum_{i}\varsigma_i\ket{w_i}\bra{v_i}$ is a singular value decomposition. 
Then there is an $m= \mathcal{O} \Big(\frac{\gamma}{\delta}
\log \left(\frac{\gamma}{\epsilon} \right)\Big)$ and an efficiently computable $\Phi\in\mathbb{R}^m$ such that
\begin{equation}
\left(\bra{+}\otimes\widetilde{\Pi}_{\leq\frac{1-\delta}{\gamma}}\right)U_\Phi \left(\ket{+}\otimes\Pi_{\leq\frac{1-\delta}{\gamma}}\right)=\sum_{i\colon\varsigma_i\leq \frac{1-\delta}{\gamma} }\tilde{\varsigma}_i\ket{w_i}\bra{v_i} , \text{ where } \Big|\!\Big|\frac{\tilde{\varsigma}_i}{\gamma\varsigma_i}-1 \Big|\!\Big|\leq \epsilon.
\end{equation}
Moreover, $U_\Phi$ can be implemented using a single ancilla qubit with $m$ uses of $U$ and $U^\dagger$, $m$ uses of C$_\Pi$NOT and $m$ uses of C$_{\widetilde{\Pi}}$NOT gates and $m$ single qubit gates.
Here,
\begin{itemize}
\item C$_\Pi$NOT$:=X \otimes \Pi + I \otimes (I - \Pi)$ and a similar definition for C$_{\widetilde{\Pi}}$NOT; see Definition 2 in \cite{gilyen2019quantum},
\item $U_\Phi$: alternating phase modulation sequence; see Definition 15 in \cite{gilyen2019quantum},
\item $\Pi_{\leq \delta}$, $\widetilde{\Pi}_{\leq \delta}$: singular value threshold projectors; see Definition 24 in \cite{gilyen2019quantum}.
\end{itemize}
\end{lemma}

\begin{lemma}[Projector]
\label{lemma: projector}
The block encoding of a projector $\ket{j-1}\bra{j-1}$ (for any $j=1,2, ...,n$) by a circuit of depth $\mathcal{O}\big( \log n \big)$ 
\end{lemma}
\noindent
\textit{Proof.} First we note that it takes a circuit of depth $\mathcal{O}(1)$ to generate $\ket{j-1}$ from $\ket{0}$. Then Lemma \ref{lemma: improveddme} can be used to construct the block encoding of $\ket{j-1}\bra{j-1}$. 
\begin{lemma}[Theorem 2 in \cite{rattew2023non}]
\label{lemma: diagonal}
     Given an n-qubit quantum state specified by a state-preparation-unitary $U$, such that $\ket{\psi}_n=U\ket{0}_n=\sum^{N-1}_{k=0}\psi_k \ket{k}_n$ (with $\psi_k \in \mathbb{C}$ and $N=2^n$), we can prepare an exact block-encoding $U_A$ of the diagonal matrix $A = {\rm diag}(\psi_0, ...,\psi_{N-1})$ with $\mathcal{O}(n)$ circuit depth and a total of $\mathcal{O}(1)$ queries to a controlled-$U$ gate  with $n+3$ ancillary qubits.
\end{lemma}

\section{Proof of Lemma \ref{lemma: constructingm1}}
\label{sec: constructingm1}
The first tool we need is an efficient construction of the so-called circulant matrix $L$, which was provided in Appendix D of \cite{nghiem2025quantum}. It turns out that there is a quantum circuit of depth $\mathcal{O}(\log n)$ which is a block encoding of a $n \times n$ circulant matrix $L$. A circulant matrix $L$ of size $n\times n$ is a special type of Toeplitz matrix, that is formally defined as:
\begin{align}
    L = \begin{pmatrix}
        l_1 & l_2 & \cdots & l_n \\
        l_n & l_1 & \cdots & l_{n-1} \\
        \vdots & \vdots & \ddots & \vdots \\
        l_2 & l_3 & \cdots & l_1
    \end{pmatrix}
\end{align}
That is, the $i$-th row is the $i-1$-th row shifted to the right by one step. The objective of Lemma \ref{lemma: constructingm1} is to construct the block encoding of $\frac{1}{\sqrt{n}} \mathcal{M}_3$, where $\mathcal{M}_3$ is defined as:
\begin{align}
    \Mcal_3=  \frac{1}{\mathcal{P}}\sum_{i=1}^n \big(f'(x_{i+1})-f'(x_i) \big)\ket{i-1}\bra{i-1} 
\end{align}
and $f'(x_{n+1}) \equiv f'(x_1)$. Recall that we have the block encoding of $\mathcal{M}_1 =  \frac{1}{\mathcal{P}} \sum_{i=1}^n  f'(x_i) \ket{i-1}\bra{i-1} $. As $H^{\otimes \log n}$ is unitary, we can use it with Lemma \ref{lemma: product} to construct the block encoding of:
\begin{align}
    \mathcal{M}_1 H^{\otimes \log n} = \frac{1}{ \sqrt{n}\mathcal{P}}\sum_{i=1}^n   f'(x_i) \ket{i-1} \bra{0} + (...)
\end{align}
where $(...)$ denotes the irrelevant part. The above operator contains $ \frac{1}{ \sqrt{n}\mathcal{P}}\sum_{i=1}^n   f'(x_i) \ket{i-1} $ in the first column. Now we choose a circulant matrix $L$ to be a $n \times n$ matrix:
\begin{align}
    L = \begin{pmatrix}
        -1 & 1 &0 & 0 \cdots &  0\\
        0 & -1 & 0 & 1 \cdots & 0\\
        \vdots & \vdots & \vdots & \ddots & \vdots \\
        1 & 0  & 0 &  \cdots & -1
    \end{pmatrix}
\end{align}
Then use Lemma \ref{lemma: product} to construct the block encoding of:
\begin{align}
    L \frac{1}{ \sqrt{n}\mathcal{P}} \sum_{i=1}^n   f'(x_i) \ket{i-1} \bra{0} + (...)   = \frac{1}{ \sqrt{n} \mathcal{P}} \sum_{i=1}^n \big( f'(x_{i+1}) - f'(x_i)  \big) \ket{i-1} + (...)
\end{align}
We use the block encoding above combined with Lemma \ref{lemma: diagonal} to construct the block encoding of $ \frac{1}{ \sqrt{n}\mathcal{P}} \sum_{i=1}^n \big( f'(x_{i+1}) - f'(x_i)  \big) \ket{i-1}\bra{i-1}$, which is exactly $\frac{1}{\sqrt{n} } \Mcal_3$.

\section{Detail of Lemma \ref{lemma: detaillambdax} }
\label{sec: detaillambdax}
This result have appeared in \cite{nghiem2025quantum} as well, so we recapitulate their procedure here. Suppose that $\ket{\Phi_1},\ket{\Phi_2}$ be the given two states that are generated by $U_1,U_2$. Consider the following state $ \frac{1}{\sqrt{2}} \ket{0} \ket{\Phi_1 } + \frac{1}{\sqrt{2}} \ket{1} \ket{\Phi_2}$, which can be generated by first creating $\frac{1}{\sqrt{2} } (\ket{0} + \ket{1}) \ket{\bf 0} $ and use $U_{1}, U_{2}$ controlled by $\ket{0},\ket{1}$ respectively to generate $\ket{\Phi_1},\ket{\Phi_2}$ entangled to corresponding register. Now we apply Hadamard gate to the first register, and append an ancilla $\ket{0}$, then we obtain $\frac{1}{2} \ket{0}\ket{0} \big( \ket{\Phi_1 }  + \ket{\Phi_2} \big) + \frac{1}{2} \ket{0}\ket{1}\big(  \ket{\Phi_1 }  - \ket{\Phi_2}\big) $. Use the second qubit as controlled bit, and apply $X$ on the first ancilla qubit, we obtain the state:
\begin{align}
    \frac{1}{2} \ket{0}\ket{0} \big( \ket{\Phi_1 }  + \ket{\Phi_2} \big) + \frac{1}{2} \ket{1}\ket{1}\big(  \ket{\Phi_1 }  - \ket{\Phi_2}\big) 
    \label{16}
\end{align}
Tracing out the second and last register, we have the following density state on the first ancilla $\rho = \frac{1}{2} \big(1 + \braket{\Phi_1,\Phi_2}\big) \ket{0}\bra{0} + \frac{1}{2} \big( 1 - \braket{\Phi_1,\Phi_2}  \big) \ket{1}\bra{1}$. We note that Lemma \ref{lemma: improveddme} allows us to block-encode $\rho$. It is trivial to obtain the block encoding of $\frac{1}{2} \ket{0}\bra{0} + \frac{1}{2}\ket{1}\bra{1}$, e.g., use Lemma \ref{lemma: scale} with scaling factor $p=2$ combined with a block encoding of $\Ibb_2 = \ket{0}\bra{0} + \ket{1}\bra{1}$, which is trivial to prepare. Then we use Lemma \ref{lemma: sumencoding} to construct the block encoding of $\rho - \frac{1}{2} \ket{0}\bra{0} - \frac{1}{2}\ket{1}\bra{1} $, which is $\frac{\braket{\Phi_1,\Phi_2}}{4} \ket{0}\bra{0 } - \frac{\braket{\Phi_1,\Phi_2}}{4} \ket{1}\bra{1} $.

Now we use $U_X$ and apply it to $\ket{\bf 0}\sum_{i=1}^n \sqrt{\lambda_i} \ket{i-1}$, we then obtain the state $\ket{\Phi_1}= \ket{\bf 0}\sum_{i=1}^n x_i \sqrt{\lambda_i} \ket{i-1} +\ket{\rm Garbage} $ (see Eqn.~\ref{eqn: action}). Defining $\ket{\Phi_2}= \sum_{i=1}^n \sqrt{\lambda_i}\ket{i-1}$, then it is straightforward to see that:
\begin{align}
    \braket{\Phi_1,\Phi_2} = \sum_{i=1}^n x_i \lambda_i 
\end{align}
Using the procedure outlined above, we then obtain the block encoding of the desired operator in Lemma \ref{lemma: detaillambdax}.

\section{Proof of Lemma \ref{lemma: multivariatefunction}}
\label{sec: multivariatefunction}
We remind that this is the result of \cite{nghiem2025quantum} so we directly quote their construction in the following. Let the coordinates of $\xbf_1$ be $(x_{1,1},x_{2,1},...,x_{M,1}) $. Similarly, coordinates of $\xbf_2$ is $(x_{1,2},x_{2,2},...,x_{M,2}) $, ..., of $\xbf_n$ is $(x_{1,n},x_{2,n},...,x_{M,n}) $. Recall from the above univariate case that we are provided (via amplitude encoding) with an efficient circuit that generates the state, or a state that contains $(x_1,x_2,...x_n)^T $ in its first $n$ entries. In this multivariate case, suppose via the same means, e.g., amplitude encoding, we are provided with a state containing $(x_{1,1},x_{1,2}, ..., x_{1,n})^T$, $(x_{2,1},x_{2,2}, ..., x_{2,n})^T$, ..., $(x_{M,1},x_{M,2}, ..., x_{M,n})^T$. Then Lemma \ref{lemma: diagonal} allows us to construct the block encoding of $ \bigoplus_{j=1}^n x_{i,j} $, for $i=1,2,...,M$. Then Lemma \ref{lemma: product} can be applied to obtain the transformation $ \bigoplus_{j=1}^n x_{i,j}  \longrightarrow   \bigoplus_{j=1}^nx_{i,j}^{k_i} $ for $i=1,2,...,M$. Then apply Lemma \ref{lemma: product} to construct the block encoding their products, which is $ \bigoplus_{j=1}^n x_{1,j}^{k_1} x_{2,j}^{k_2}... x_{M,j}^{k_M} $. Then we use Lemma \ref{lemma: scale} to insert the factor $a_j$, i.e., we obtain the block encoding of  $ \bigoplus_{j=1}^n  a_k x_{1,j}^{k_1} x_{2,j}^{k_2}... x_{M,j}^{k_M} $. Finally, we reppeat the above procedure to construct the same block encoding but for different $(k_1,k_2,...,k_M)$, then we use Lemma \ref{lemma: sumencoding} to construct the block encoding of $ \frac{1}{K} \bigoplus_{j=1}^n \sum_{k=1}^K a_k x_{1,j}^{k_1} x_{2,j}^{k_2}... x_{M,j}^{k_M}  = \frac{1}{K} \bigoplus_{j=1}^n f(\xbf_j)  $

\end{document}